# ENERGY EFFICIENT FULL ADDER CELL DESIGN WITH USING CARBON NANOTUBE FIELD EFFECT TRANSISTORS IN 32 NANOMETER TECHNOLOGY

Ali Ghorbani and Ghazaleh Ghorbani

Young Researchers and elite Club, Meymeh Branch,
Islamic Azad University, Meymeh, Iran

*ABSTRACT*

*Full Adder is one of the critical parts of logical and arithmetic units. So, presenting a low power full adder cell reduces the power consumption of the entire circuit. Also, using Nano-scale transistors, because of their unique characteristics will save energy consumption and decrease the chip area. In this paper we presented a low power full adder cell by using carbon nanotube field effect transistors (CNTFETs). Simulation results were carried out using HSPICE based on the CNTFET model in 32 nanometer technology in Different values of temperature and VDD.*

## 1. INTRODUCTION

Full adder cells are one of the most important parts of the arithmetic operations [1]. So, using an optimized full adder cell increase the performance of these operations. So, designing faster and low-power FAs was the driving force behind many results reported during the last decade. The main purpose of such designs has create faster FAs while also reducing their power consumption.

In this paper we present a low power full adder cell.

Moore believes that the number of transistors on a chip will be double about every two year [2].To achieve this aim, we have to reduce the size of the transistor into Nano-scale region [3-4]. Due to the limitations of silicon based Field Effect Transistors (FETs) we need to use an efficient alternative technology.

Carbon Nanotube Field Effect Transistor (CNTFET) is an optimized alternative for the conventional CMOS technology [4].

This paper organize as follows:

 In section 2, we review the carbon nanotube field effect transistors (CNTFETs) and their Specifications, The previous designs of full adders discuss in section 3 and then present our proposed design in section 4 and finally, simulation results in section 5. Conclusion will present in Section 6.

## 2. CNTFETs

Carbon Nano Tubes (CNTs) are cylindrical shape of graphite sheets that rolled as tube [5]. The carbon nanotubes have chirality vector (m, n) that depend on values of them and either they can be metallic or semiconducting [6-7].

Carbon nanotube field effect transistors (CNTFETs) are a kind of transistors that use carbon nanotubes as their channel. There are two form of CNTFET: Schottky Barrier CNTFET (SB-CNTFET) and MOSFET-like CNTFET. In MOSFET-like source and drain are made of doped carbon nanotubes and the intrinsic semiconducting carbon nanotubes are used in the channel region. The channel in the Schottky barrier (SB) CNTFET is an intrinsic semiconducting carbon nanotube and direct contacts of the metal with the semiconducting nanotubes made for source and drain regions [8-9].

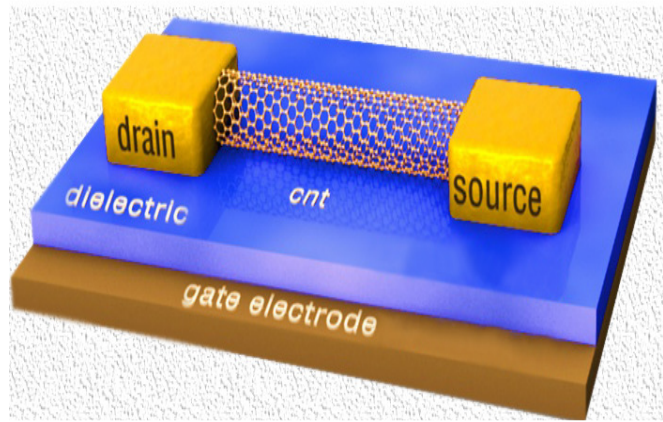

Figure 1. Schematic diagram of a CNTFET

CNFETs are one of the molecular devices that avoid most fundamental silicon transistor restriction and have ballistic or near ballistic transport in their channel [10]. Therefore a semiconductor carbon nanotube is appropriate for using as channel of field effect transistors [7]. Applied voltage to the gate can be control the electrical conductance of the CNT by changing electron density in the channel.

The diameter of carbon nanotube can be expressed as equation 1 [11]:

$$(1) \quad D_{CNT} = \frac{a\sqrt{n^2 + m + m^2}}{\pi}$$

The threshold voltage has reversed relationship between nanotube diameters that shows in equation 2.

$$(2)\quad V_{TH} = \frac{0.42}{d(nm)} V$$

So, by changing the diameter of the nanotube threshold voltage can be define and also, we can define the behaviour of the transistors.

## 3. PREVIOUS DESIGNS

Many full adder cells have presented in the past [1, 5, 12-19]. In [1, 5, and 12] full adders were designed based on XOR/XNOR circuits. The capability of majority function encourage the designers to use this function in their designs (figure 2) [13-18]. The majority function show in equation3.

$$(3)\quad MAJORITY(A, B, C) = AB + BC + AC = Cout$$

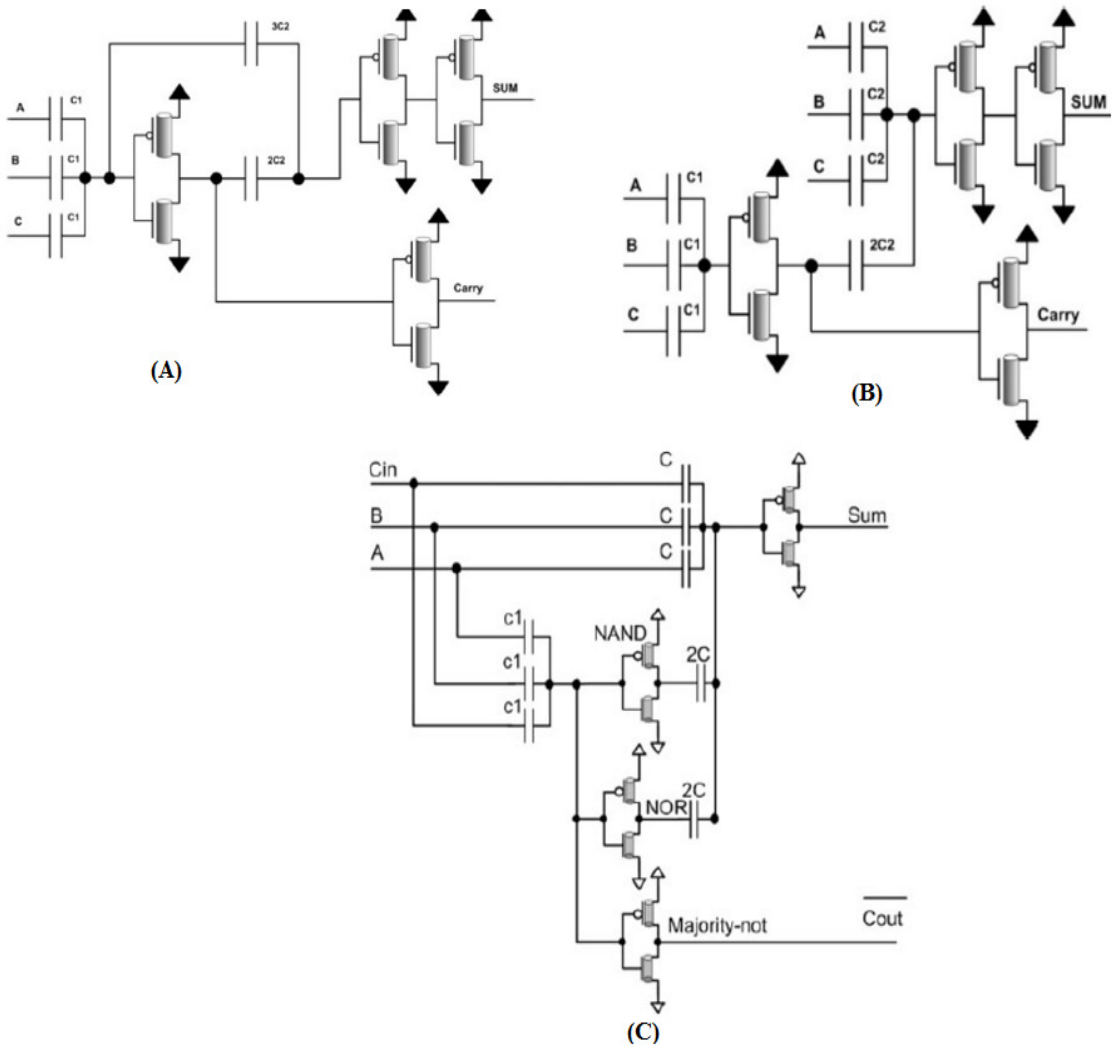

Figure 2. Full adder designs with majority function

Using the capacitor and resistor elements affect the power of design. In [19] the full adder cell designed without any capacitors and resistors. In this paper we present a new CNTFET full adder design with reducing the number of transistors and also we don't use any capacitor or resistor. (Figure 3)

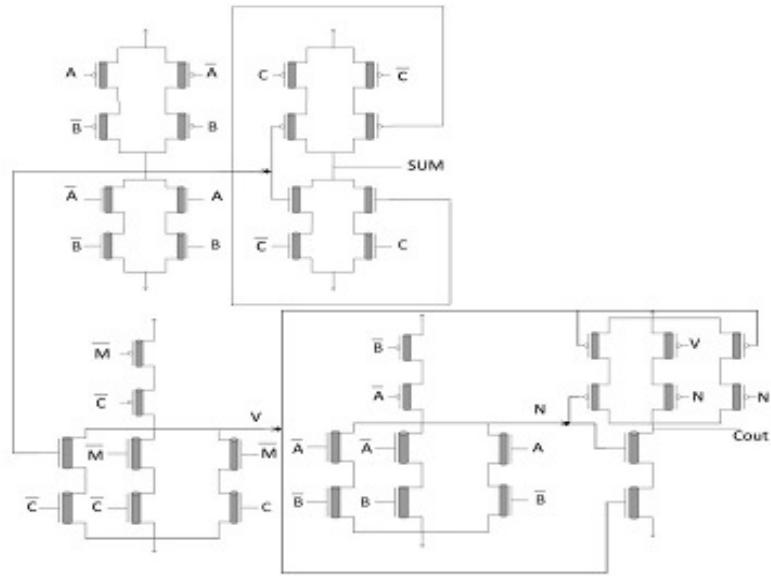

Figure 3. Full adder 4

## 4. NEW CNTFET FULL ADDER DESIGN

In this paper we present a full adder cell design with using carbon nanotube field effect transistors in 32 nm technology. This design implements with 24 carbon nanotube field effect transistors (CNTFETs) and due to the implementation of this circuit in Nano-scale, the power of this design is better than the CMOS one. Also this circuit implemented by reducing the number of transistors in comparison with the previous CNTFET designs. This factor is also optimizes power supply of the circuit. This design implements based on equation 4.Schematic of this circuit design shows in figure 4.

$$\text{Cout} = AB + BC + AC$$
$$SUM = ABC + A\overline{B}\overline{C} + \overline{A}B\overline{C} + \overline{A}\overline{B}C$$
(4)
$$= ABC + A(\overline{AB} + \overline{AC} + \overline{BC}) + B(\overline{AB} + \overline{AC} + \overline{BC}) + C(\overline{AB} + \overline{AC} + \overline{BC})$$
$$= ABC + A\overline{(AB + AC + BC)} + B\overline{(AB + AC + BC)} + C\overline{(AB + AC + BC)}$$
$$= ABC + A(\overline{\text{Cout}}) + B(\overline{\text{Cout}}) + C(\overline{\text{Cout}})$$

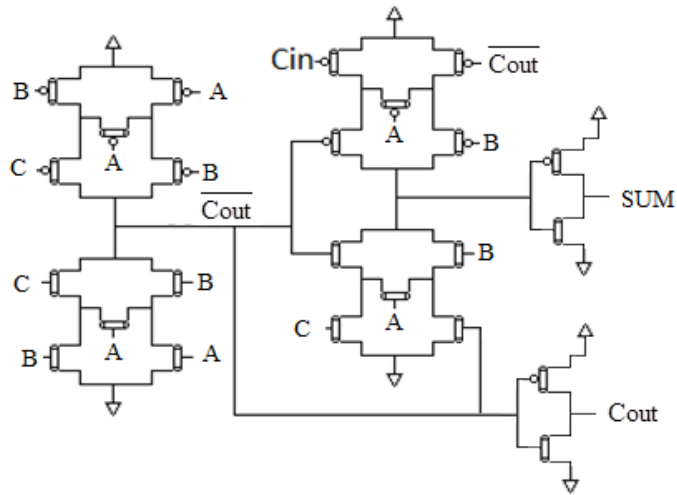

Figure 4. New CNTFET Full adder design

## 5. SIMULATION RESULTS

The simulation results show in tables 1-3. These Simulation results were carried out using HSPICE based on the CNTFET model in 32 nanometer technology. Tables1-3 show the Power, Delay and PDP results in Different values of temperature and VDD. As shown in table1 and 3, the Power and Power-Delay Product (PDP) of the circuit have increased by growing the VDD but it has not appreciably changes by growing the temperature (figure 5-6). The Delay parameter has decreases by increasing the VDD (as shown in table 2).

Table 1: Power consumption results for various amounts of VDD and Temperature (*10-9)

| $V_{DD}$\Temp ℃ | 0 | 9 | 18 | 27 | 36 | 45 | 54 |
|---|---|---|---|---|---|---|---|
| 0.7 | 2.6936 | 2.6913 | 2.7776 | 2.6864 | 2.72 | 2.6759 | 2.6842 |
| 0.8 | 3.8247 | 3.7974 | 3.7915 | 3.8371 | 3.8419 | 3.8521 | 3.8769 |
| 0.9 | 5.8277 | 5.834 | 5.8614 | 5.9341 | 5.842 | 5.9107 | 5.8542 |
| 1 | 8.5178 | 8.5429 | 8.5635 | 8.6274 | 8.9283 | 8.6518 | 8.6511 |
| 1.1 | 12.505 | 12.266 | 12.264 | 12.313 | 12.927 | 12.26 | 12.565 |
| 1.2 | 18.692 | 19.459 | 18.808 | 18.326 | 18.504 | 19.466 | 18.645 |

Table 2: Delay results for various amounts of VDD and Temperature(*10-10)

| $V_{DD}$\Temp ℃ | 0 | 9 | 18 | 27 | 36 | 45 | 54 |
|---|---|---|---|---|---|---|---|
| 0.7 | 1.3229 | 1.3186 | 1.3138 | 1.3077 | 1.3032 | 1.2981 | 1.2944 |
| 0.8 | 1.2403 | 1.2342 | 1.2301 | 1.2258 | 1.2208 | 1.2175 | 1.2132 |
| 0.9 | 1.2004 | 1.1961 | 1.1938 | 1.1898 | 1.1851 | 1.1802 | 1.1775 |
| 1 | 1.1385 | 1.134 | 1.1298 | 1.1235 | 1.1187 | 1.1142 | 1.1101 |
| 1.1 | 1.1119 | 1.1087 | 1.1021 | 1.0969 | 1.0917 | 1.0892 | 1.0855 |
| 1.2 | 1.0921 | 1.0891 | 1.083 | 1.0788 | 1.0735 | 1.0701 | 1.0671 |

Table 3: PDP results for various amounts of VDD and Temperature(*10-19)

| $V_{DD}$\Temp ℃ | 0 | 9 | 18 | 27 | 36 | 45 | 54 |
|---|---|---|---|---|---|---|---|
| 0.7 | 3.5634 | 3.5487 | 3.6492 | 3.5130 | 3.5447 | 3.4736 | 3.4744 |
| 0.8 | 4.7438 | 4.6868 | 4.6639 | 4.7035 | 4.6902 | 4.6899 | 4.7035 |
| 0.9 | 6.9956 | 6.9780 | 6.9973 | 7.0604 | 6.9234 | 6.9758 | 6.8933 |
| 1.0 | 9.6975 | 9.6876 | 9.6750 | 9.6929 | 9.9881 | 9.6398 | 9.6036 |
| 1.1 | 13.9043 | 13.5993 | 13.5162 | 13.5061 | 14.1124 | 13.3536 | 13.6393 |
| 1.2 | 20.4135 | 21.1928 | 20.3691 | 19.7701 | 19.8640 | 20.8306 | 19.8961 |

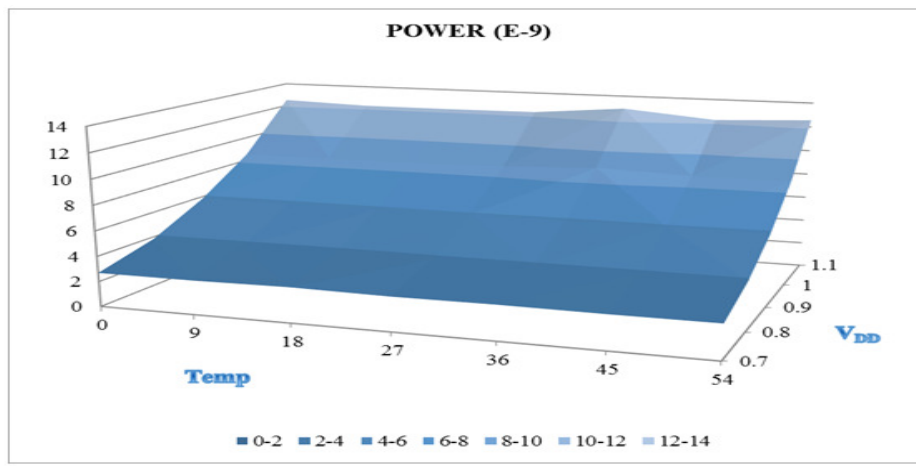

Figure 5. Chart of Power consumption results for various amounts of VDD and Temperature

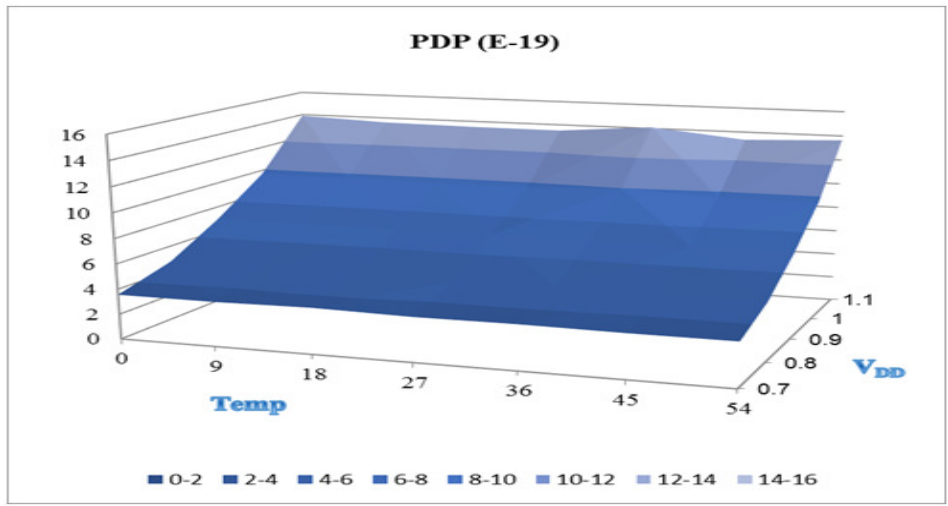

Figure 6. Chart of PDP results for various amounts of VDD and Temperature

Finally we compare the simulation results of this Full Adder cell design with previous designs in 0.65 V VDD (Table 4).

Table4. Comparing the results with previous designs

|  | Power | Delay | PDP |
| --- | --- | --- | --- |
| Full adder-1[14] | 5.23E-07 | 7.97E-11 | 4.17E-17 |
| Full adder-2[16] | 4.71E-07 | 8.82E-11 | 4.15E-17 |
| Full adder-3[17] | 7.12E-07 | 7.51E-11 | 5.35E-17 |
| Full adder-4[19] | 1.35E-08 | 3.45E-11 | 4.663E-19 |
| **New FA design** | **2.66E-09** | **1.30E-10** | **3.513E-19** |

## 6. CONCLUSION

In this paper we presented a low power full adder cell with using carbon nanotube field effect transistors. We used lower number of transistors and reduced the chip area by using these Nano-scale transistors. Results of section 5 carried out using HSPICE based on the CNTFET model in 32 nanometer technology in Different values of temperature and VDD. We compared these simulation results with previous similar works and showed that we achieve an optimized power saving.

## ACKNOWLEDGEMENTS

This paper is extracted from a research project with the title "A new binary full adder cell by using carbon Nanotube field effect transistors".

## REFERENCES

[1] Bui H T, Wang Y, and Jiang. Design and analysis of low-power 10-transistor full adders using XOR–XNOR gates. Circuits and Systems II: Analog and Digital Signal Processing, IEEE Transactions. Jan.2002; 49, 1: 25–30.

[2] Moore G E. Progress in digital integrated electronics. In: IEEE. Retrieved 2011 Electron Devices Meeting 1975; 11–27.

[3] Wu J, Shen Y, Reinhardt K, Szu H, Dong B. A Nanotechnology Enhancement to Moore's Law. Applied Computational Intelligence and Soft Computing; 2013: Article ID 426962, 13 pages, 2013. doi:10.1155/2013/426962.

[4] Appenzeller J. Carbon Nanotubes for High-Performance Electronics Progress and Prospect. Proceedings of the IEEE. Feb 2008; 96, 2, 201-211.

[5] Lin J, Hwang Y. A Novel High-Speed and Energy Efficient 10-Transistor Full Adder Design. IEEE Transactions on Circuits and Systems. May 2007; 54, 5, May 2007: 1050-1059.

[6] Jiang Y, Al-Sheraidah A, Wang Y,Sha E, Chung J. A novel multiplexer-based low power full-Adder. Circuits and Systems II: Express Briefs, IEEE Transactions; 51, 7: 345 – 348.

[7] Raychowdhury A, Roy K. Carbon Nanotube Electronics: Design of High-Performance and Low-Power Digital Circuits. Circuits and Systems I: Regular Papers, IEEE Transactions, Nov 2007; 54, 11:2391-2401.

[8] Abdolahzadegan SH, Keshavarzian P, Navi K. MVL Current Mode Circuit Design Through Carbon Nanotube Technology. European Journal of Scientific Research.2010; 42, 1:152-163.

[9] Javey A, Guo J, Wang Q, Lundstrom M, Dai H. Ballistic carbon nanotube field-effect transistor, Nature,2003; 424:654-657.

[10] Raychowdhury A, Roy K. Carbon-Nanotube-Based Voltage-Mode Multiple-Valued Logic Design. IEEE Trans. Nanotechnology, March 2005; 4, 2:168-179.

[11] McEuen P L, Fuhrer M S, Park H. Single Walled Carbon Nanotube Electronics. Nanotechnology, IEEE Transactions, 2002; 1, 1:78-85.

[12] Goel S, Kumar A, Bayoumi M A. Design of robust, energyefficient full adders for deep submicrometer design using hybrid-CMOS logic style. IEEE Trans on VLSI Systems, 2006; 14, 12:1309-1321.

[13] Navi K, Maeen M, Foroutan V, Timarchi S, Kavehei O. A novel low power full-adder cell for low voltage. INTEGRATION, the VLSI Journal, 2009; 42:457-467.

[14] Navi K, Momeni A, Sharifi F, Keshavarzian P. Two novel ultra high speed carbon nanotube Full-Adder cells. IEICE Electronics Express, 2009; 6, 19:1395-1401.

[15] Navi K, Foroutan V, Rahimi Azghadi M, Maeen M, Ebrahimpour, Kaveh M, Kavehei O. A Novel Ultra Low-Power Full Adder Cell with New Technique in Designing Logical Gates Based on Static CMOS Inverter. Microelectronics Journal, Elsevier, 2009; 40:1441-1448.

[16] Navi K, Sharifi Rad R, Moaiyeri M H, Momeni A. A Low-Voltage and Energy-efficient Full Adder Cell Based on Carbon Nanotube Technology. Nano Micro Letters, 2010; 2,2:114-120.

[17] Navi K, Rashtian M, Khatir A, Keshavarzian P, Hashemipour O. High Speed Capacitor-Inverter Based Carbon Nanotube Full Adder. Nanoscale Ress Lett, 2010; 5:859-862.

[18] Khatir A, Abdolahzadegan SH, Mahmoudi I.High Speed Multiple Valued Logic Full Adder Using Carbon Nano Tube Field Effect Transistor. International Journal of VLSI design & Communication Systems (VLSICS), 2011; 2, 1.

[19] Ghorbani A, Sarkhosh M, Fayyazi E, Mahmoudi N, Keshavarzian P.A Novel Full Adder Cell Based On Carbon Nanotube Field Effect Transistor. International Journal of VLSI design & Communication Systems (VLSICS), 2012; 3, 3.